# A NOVEL SOLUTION TO THE SHORT RANGE BLUETOOTH COMMUNICATION

Preetha K G

Department of Information Technology,
Rajagiri School of Engineering & Technology, Cochin, India
preetha_kg@rajagiritech.ac.in

## ABSTRACT

*Bluetooth is developed for short range communication. Bluetooth Devices are normally having low power and low cost. This is a wireless communication technology designed to connect phones, laptops and PDAs. The greater availability of portable devices with Bluetooth connectivity imposes wireless connection between enabled devices. On an average the range of Bluetooth devices is about 10 meters. The basic limitation of the Bluetooth communication is this range limitation. In this paper I have studied the limitations of Bluetooth communication and consider range constraint as the major limitation. I propose a new expanded Blue tooth network to overcome the range constraint of Bluetooth device. This creates a network of Bluetooth enabled devices that will include laptops, set top devices and also mobile phones. The main purpose of this proposal is to establish a network will enable the users to communicate outside the range without any range constraint.*

## KEYWORDS

*Bluetooth, Pico net, Scatter net, Frequency hoping, ISM*

## 1. INTRODUCTION

Mobility originated from the desire to move either toward resources or away from scarcity. Mobile computing is concerned about the movement of physical devices, user applications and mobile agents. Bluetooth technology developed by Bluetooth Special interested Group and the main purpose of this technology is to replace the connection cables used to connect devices, with one universal short-range radio link. Radio waves are used as the communication medium. These radio waves are operating in the unlicensed ISM band with 2.45 GHz frequency. Bluetooth uses a spread-spectrum frequency-hopping technique which takes a narrowband signal and spreads it over a broader portion of the available radio frequency band. In frequency hoping spread spectrum (FHSS), the total available frequency is split into many channels of smaller bandwidth. Transmitter and receiver stay on one of these channels for a certain period and then hop to another channel. The channel usage pattern is called hopping pattern. There is a frequency selection module gives the frequency depending on masters clock. The hopping rate is 1600 hopes/sec. There are two types of FHSS scheme. Slow hopping and fast hopping. In slow hopping the transmitter uses one frequency for several bit transmissions. In fast hopping the transmitter changes the frequency many times during the transmission of single bit. This method reduces the interference with other devices. Since only the intended receiver is aware of the transmitter's hopping pattern, only that receiver can make sense of the data being transmitted. This technique ensures Bluetooth's security and limits interference. Bluetooth is hoped to be a very low cost type of networking, and, as it becomes more widespread, the cost of adding Bluetooth to devices should drop down to perhaps no more than $5-10. Bluetooth is designed to be compatible across a range of very different operating systems and devices.

The main intention of Bluetooth technology is to eliminate cables, connectors for connection between devices. So the expenses for cables and connectors are reduced, which gives large economic benefit [1]. The interesting characteristics of Bluetooth devices are very small in size low power and low cost. The devices are small in size so it is portable and it can be attached to any device. Bluetooth devices are low cost devices and the utilization of power is also very less. The effective range of Bluetooth devices is 32 feet (10 meters). A major challenge lies in Bluetooth communication is this range constraint. This paper studies the range limitation of Bluetooth communication and also proposes a solution to this range constraint. The paper tries to expand the r as much as possible by exploring the basic routing algorithm.

The rest of this paper is organized as follows. The overview of Bluetooth technology is given in Section 2. Section 3 describes the complexities involved in Bluetooth communication. A brief description of the proposed protocol is given in section 4. Salient features of new protocol are described in section in 5. The application and the area of significance are discussed in section 6 and conclusion is given in section 7.

## 2. AN OVERVIEW OF BLUETOOTH ARCHITECTURE

The main motive of developing Bluetooth is low cost device for short range (10 meter) communication. It places a vital roll in wireless communication. The goal of Bluetooth specification is the uniform structure for a wide range of devices to connect and communicate with each other. These specifications are developed and licensed by the Bluetooth Special Interest Group (SIG). The Bluetooth SIG consists of companies in the areas of telecommunication, computing, networking, and consumer electronics. The data transfer rate of Bluetooth is 1Mbps. The Bluetooth specification can support three synchronous voice channels at 64 Kbps each. Bluetooth devices typically require 1mW of power to operate.

The basic unit of Bluetooth device is piconet. The Bluetooth devices are organized as a network of two to eight Pico nets. The architecture of a Bluetooth piconet is master-slave architecture and it consists of a single master device and one or more slave devices. Each node in the Pico net uses the same frequency hopping technique. A slave or master in one Pico net can communicate with the master or slave in other Pico net. A common node of two piconets is called a bridge node. This bridging structure is called scatter net (Figure 1). In the figure, M represent master and s represents slave. with this the number of nodes can be expanded up to 255 in the network. But only one master is allowed in one piconet.

The structure of Bluetooth is Master Slave and operation is packet based . One master may communicate with up to 7 slaves in a piconet; all devices share the master's clock. Packet exchange is based on the basic clock, defined by the master, which ticks at 312.5 μs intervals. Two clock ticks make up a slot of 625 μs; two slots make up a slot pair of 1250 μs. In the simple case of single-slot packets the master transmits in even slots and receives in odd slots; the slave, conversely, receives in even slots and transmits in odd slots. Packets may be 1, 3 or 5 slots long but in all cases the master transmit in even slots and the slave transmit in odd slots.

The working of Bluetooth network is depends on Bluetooth protocol stack. Bluetooth network is divided into different layers as shown in (Figure 2). Each layer has its own protocol. This reference model is different from other existing reference model. The Bluetooth radio layer is the lowest layer similar to the physical layer of internet model. It defines the requirements of the Bluetooth transceiver device operating in the 2.4GHz ISM band. Frequency hoping technique is used to reduce the interferences. The basic radio in this is a hybrid spread spectrum radio. Typically, the radio operates in a frequency-hopping manner in which the 2.4GHz ISM band is broken into 79 1MHz channels that the radio randomly hops through while transmitting and receiving data. Base band layer is similar to Mac sub layer in internet. It manages physical channels and links. TDMA is used as the access method. Single slave and multiple slave communication is allowed in this layer. There are two links used for communication. Synchronous connection oriented and asynchronous connection less. Next layer is the Logical Link Control and Adaptation Protocol (L2CAP). It is the basis for the cable replacement usage of Bluetooth. It supports higher level protocol multiplexing, packet segmentation and reassembly, and the conveying of quality of service information. The Link Manager Protocol (LMP) is used by the Link Managers for link set-up and control. The Host Controller Interface (HCI) provides a command interface to the Base band Link Controller and Link Manager, and access to hardware status and control registers. The RFCOMM protocol provides emulation of serial ports over the L2CAPprotocol. The protocol is based on the ETSI standard TS 07.10. The Service Discovery Protocol (SDP) provides a mechanism to discover which services are provided by or available through a Bluetooth device. It also allows applications to determine the characteristics of those available services. LMP, L2CAP, SDP are called Bluetooth core protocols.

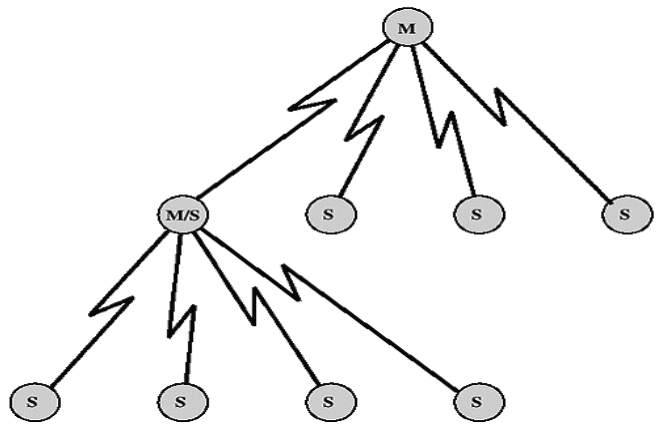

Figure 1.Scatternet

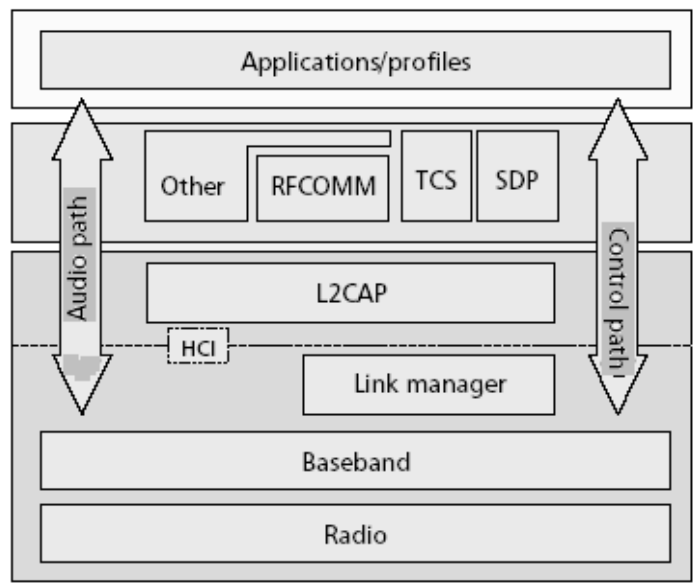

Figure 2.Bluetooth Layers

Bluetooth is the major part of personal area network (PAN) [10]. Scatternet enables the formation of PAN. Bluetooth networking transmits data via low-power radio waves. It communicates on an International agreed frequency (ISM) of 2.45 gigahertz. A Time Division Multiplexing (TDM) technique is used in this communication. This divides the channel into 625p slots and, with a 1 MBPS symbol rate, a slot can carry up to 625 bits. Transmission occurs in packets and each packet is transmitted on a different hop frequency with a maximum frequency hopping rate of 1600 hops/s. Based on output power rating Bluetooth devices are divided into three classes. Most powerful devices are in class 1 category. This is defined for defined for long range. These can have up to 100 mW of power, and a range of about 40 m - 100 m (130 - 330 ft). Class 2 devices are lower power, with up to 2.5 mW of power and a range of about 15 m - 30 m (50 - 100 ft).This is defined for ordinary range. Third category is class 3defined for short range.. This uses less power up to 1 mW and a range of about 5 m - 10 m (16 - 33 ft). Most of the Bluetooth devices will fall under class 2 or class 3.

## 3. ISSUES AND DEMANDS IN BLUETOOTH COMMUNICATION

The present scenario using Bluetooth has many obscurities. The main limitation is data transmission is possible only to short distances. If one path is busy with network overloading then we have to wait till the path is cleared. If the network is over loaded then the data transmission rate will be low. Other difficult problem arises in Bluetooth communication is related to its low power. The low power devices limit the range of a Bluetooth device to about 10 meters (32 feet paper). Even with the low power, Bluetooth doesn't require line of sight between communicating devices. The states of the Pico net nodes are of categories. These are active, idle, parked and sniffing. Data exchange takes place only between active nodes but the nodes periodically change its states. This imposes a greater challenge in Bluetooth design. Which node is selected as a master, and how many nodes that can be used to connect to other Pico net are the challenging areas.

Security is one of the major issues in Bluetooth networking. As Bluetooth works in unlicensed frequency band, it is more vulnerable to many security attacks. Some of the security measures in piconet are key establishment, authentication, and encryption protocols to make communication within a piconet secure using key management. This is provided by Bluetooth Specification itself. Protection of communication within the scatternet is an open challenge.

Currently, in the Bluetooth network the data rate is about 3 megabits (375 kilobytes) per second. This slow rate is only apparent when very large files are being sent. All Bluetooth enabled devices operate within the 2.4 GHz band, which is the same unlicensed frequency used by many other electronic devices such as microwaves, cordless phones, and the majority of Wi-Fi devices. This frequency sharing could lead to slower overall network performance because some of the signals collide, don't reach their destination, and have to be re-sent.

Bluetooth is primarily intended to facilitate short range data transfer using low power. The main disadvantage of this technology is the nodes involve in the data transfer have to be very near each other. Even if the device is on, due to this range constraints we can not reach the devices properly. This is represented in figure 3. This resulting most of the time the greater number of Bluetooth enabled devices are inactive. This will decrease the communication throughput. Here I am considering the range constraint as the most prominent and basic limitation of Bluetooth communication. So In this paper we are concentrating on the range limitation of Bluetooth communication and we are trying to expand the range by using the enabled intermediate devices.

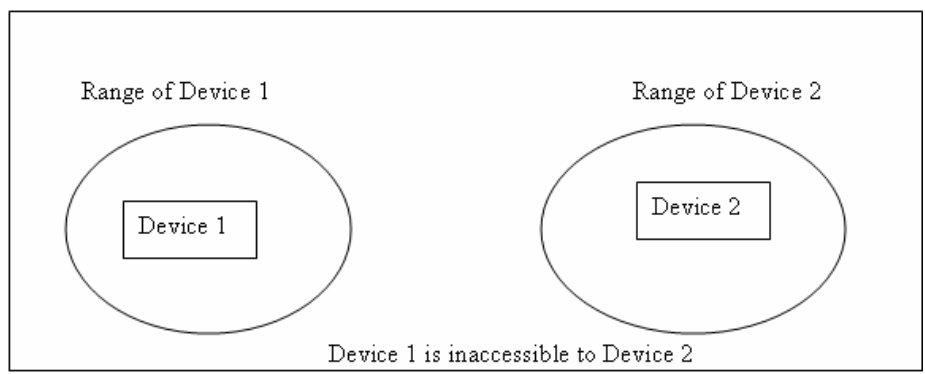

Figure 3.Range of Bluetooth devices

## 4. RANGE EXPANSION-A NEW THOUGHT

The enhancement of portable and wireless e increases the Bluetooth network applications. Here I am considering range is the major parameter for any Bluetooth application. Even if the Bluetooth device is on, many of the time the network establishment is not possible due to the out of range problem. The main goal of this paper is to overcome the basic limitation of Bluetooth devices. The range of Bluetooth devices are usually of 10 meters. This range restriction limits the application areas of Bluetooth network. This limitation is overcome by establish a network through enabled intermediary device. This allowing a device to another device outside its range through an intermediary device, which has access to both the devices (figure 4).

The transmission of a mobile host is received by all hosts within its transmission range due to the broadcast nature of wireless communication and omni-directional antennae. If two wireless hosts are out of their transmission ranges in the ad hoc networks, other mobile hosts located between them can forward their messages, which will builds a network among the mobile hosts in the area. Due to the mobility of wireless hosts, each host needs to be equipped with the capability of an autonomous system, or a routing function without any statically established infrastructure or centralized administration. The mobile hosts can move arbitrarily and can be turned on or off without notifying other hosts. The mobility and autonomy introduces a dynamic topology of the networks not only because end-hosts are transient but also because intermediate hosts on a communication path are transient.

Initially we need to establish an ad-hoc network with all available devices. It can be thought of these connections as spontaneous networks, available to who ever is in a given area. An ad hoc network is one where there are no access points passing information between participants. This network is created only for specific applications and temporary usage. In this network each node is maintaining a routing table. The routing processes are of two types static and dynamic. In static up-to date routing information is maintained in each node and it is independent of the requirement. In dynamic, route discovery process is done only when it is needed.  Whenever they want to send data then they have to find the route and create their routing table. Due to the mobility of wireless hosts, each host needs to be equipped with the capability of an autonomous system, or a routing function without any statically established infrastructure or centralized administration. The mobile hosts can move arbitrarily and can be turned on or off without notifying other hosts. The mobility and autonomy introduces a dynamic topology of the networks not only because end-hosts are transient but also because intermediate hosts on a communication path are transient.

To set up the network each device broadcast a list of accessible devices within its range. Each device updates their table of accessible devices according to this list. Each device prepared a list of other devices that can be accessed directly or indirectly. If a device wishes to send a message to another device in this list, a path is found through intermediary device through which the destination can be reached and forwards the message to the first device in the path. But the intermediate devices dynamically choose the path through which the message is forwarded depending upon the traffic and availability of devices.

In this proposal I tried to form a Mobile ad-hoc network (MANET) by using enabled Bluetooth devices. MANET consists of mobile devices forms a network. In MANETs all nodes are considered as source or router and the control of the network is distributed among nodes. A major challenge lies in MANET communication is the unlimited mobility and more frequent failures. Because of frequent short lived disconnections, the increased chances for collisions, transmission errors and the probability of missing data is more in MANETs.

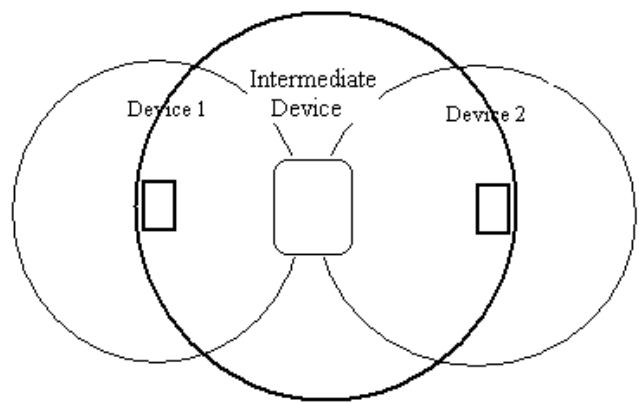

Figure 4. Expanding range using intermediate device

## 5. FEATURES OF AN EXPANDED NETWORK

In mobile ad-hoc scenario all the devices are in motion. Even if they are moving our aim is to design an efficient and stable networking system, in which, it is possible to track them and transfer data to them without much hindrances and disturbances. The path to be selected for transferring data would depend on shortest path criteria and load balancing criteria.

This paper tries to expand the range of Bluetooth data transfer by involving intermediate devices between the sender and receiver. A message from the source goes to one or more intermediate device finally ends up at destination. A number of paths from the source to destination is obtained. The number of nodes in each path is analysed from the routing table and select the path with lesser number of nodes. Once the data is received then the destination send acknowledgement. The data is going through different intermediary nodes but these devices can not recognize the data because the data is encrypted. Only the destination can decrypt the message. This is typically considers as a client server architecture. The device which sends the data is the client and receives the data is the server. The client node expands its network by searching for the Bluetooth enabled device in its range. All devices continue this searching for devices within the range until the destination is reached. A model of the proposed system is in the figure 5.

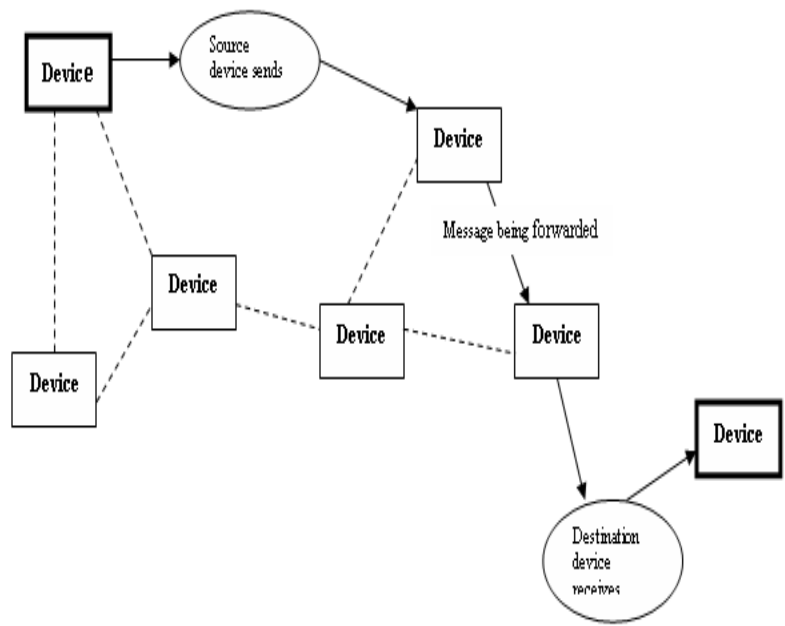

Figure 5.Model for the proposed system

This network consists of devices of smaller speed and less power. This network considers smaller network. First step of the process is to find the path between the source and the destination. The routing algorithm I have chosen is distance vector routing algorithm. This is also called Bellman-Ford or Ford-Fulkerson algorithm. This is a static algorithm and very well suited in wired network. In the wired network all the nodes are fixed. So it is easy to find the route between any two nodes. But in the case of mobile networks this will not work as such. Slight modification is needed. In the static algorithm each node finds its neighbour and send this information to all nodes. But this algorithm is not suitable for mobile network because in the mobile network all the nodes are mobile nodes. I have modified the distance vector algorithm which is used in the new Bluetooth network. The operation of the algorithm is as follows. When a node starts then it can directly access its immediate neighbours. Each node creates a list of nodes that can be accessible. Each node, on a regular basis, sends to each neighbour its own current idea of the total cost to get to all the destinations it knows of. Cost is determined by the number of nodes in the path. The neighbouring nodes examine this information and update their routing table accordingly. Over time, all the nodes in the network will discover the best next hop for all destinations, and the best total cost.

A node wants to send message to another node in the network first it check whether this node is in the range of the sender. If so then it can directly send message. Otherwise it will set a path to the destination through the intermediate devices. Each device sends their accessible device to its neighbours so the sender can calculate the shortest path to the destination. Once the routing path is finalised then sender node can access the destination. Each intermediate node in the path is involved in the routing process and each will be aware of the data transmission.

In any Bluetooth data transfer the nodes are not fixed so any node can move from the network and new node can come up at any time. If any node is added to the network then it finds its immediate neighbours and prepares its routing table. And this routing table is send to all nodes that can be directly accessible. If any node wishes to move out of the network then it send a withdraw message to their immediate neighbour nodes. In both cases all nodes update their routing table accordingly. If any node updates its routing table then inform their neighbours and send the routing table to them.

When a node send a packet to another node in the network if it reaches correctly at the destination then it will send the acknowledgement. If the sender does not get the acknowledgement before the timer turned off then the route discovery process is repeated. Sender must send the same packet through another shortest route if the current route does not exist. The path selection is crucial and is to be selected depends on the shortest path criteria and load balancing criteria.

## 6. APPLICATIONS AND AREAS OF SIGNIFICANTS

With the development of Bluetooth technology, many Bluetooth devices come into our living, such as Bluetooth earphone, Bluetooth home-network, Bluetooth head phone etc. Recently, the Bluetooth technology is the fastest growing technology which enables devices to connect and communicate. Data dissemination is the main application intended to the Bluetooth network. We can send text messages as well as picture messages to any Bluetooth enabled devices via Bluetooth communication. Bluetooth is actually the replacement of traditional wired serial communication in test equipment, GPS receivers and medical equipment. Bluetooth places an important role in wireless networking with devices where little bandwidth is needed.

The popular use of Bluetooth technology is wireless control and communication between any devices with Bluetooth capability. The devices can be cell phone, mouse, keyboard, cordless headset, camera, PDA, printer, computer etc. Bluetooth can also help different devices to communicate with each other. For example, if you have a phone, a PDA, and a computer and all the three devices have Bluetooth capabilities, then with the support of appropriate software on each device you can look up a phone number on your PDA and then place a call direct from the laptop or PDA without touching your cell phone. Ad hoc networking and remote control are the significant applications. Another attractive application is wireless networking between PCs in a confined space where little bandwidth is required. By using Bluetooth communication technology transfer of files between devices via OBEX is possible. This technology can be used in developing mobile ad-hoc networks. The various applications for this includes military applications, collaborative computing, emergency reuse, and multi- hop cellular networks etc.

## 7. CONCLUSIONS

More people fascinate for the wireless devices. Bluetooth technology is a global standard for wireless connectivity. This facilitates the replacement of wires or cables used to interconnect between devices. Bluetooth devices are small in size, low cost devices, and the utilization of power is very low. Bluetooth can imitate a universal bridge to attach the existing data networks, and also as a mechanism for forming ad-hoc networks. This is designed to operate in noisy frequency environments, the Bluetooth radio uses a fast acknowledgement and frequency hopping scheme to make the link robust. In this paper, I have listed out the common issues of Bluetooth data transmission. The major shortcoming I have found is the range limitation. So in this work I was concentrating on this range limitation. Efficiency of connection establishment has been analysed and suggested a method to overcome the basic limitation of Bluetooth communication that is the range constraint.

In this work a new network is established within the Bluetooth enabled devices. With this new network, the range of the devices that can be accessible is expanded. This expansion is done through the enabled intermediate devices. The algorithm chosen for expansion is distance vector algorithm. Normal Bellman Ford algorithm is static. But in this case the route is discovered only when it is needed. Each time every node finds its neighbouring nodes and updates their routing table accordingly. When a device tries to connect to other devices, first it finds the devices that can be accessed directly or indirectly. Then it can establish a path to the destination through the intermediate devices and forward the message. There is one drawback identified in the new network is the network overhead. Whenever the data to be send then each time the node has to find the path to the destination. During this path discovery and updating of routing table, so many packets have been sent though the network. The result is the network overhead.

**Authors**

Mrs. Preetha K G completed her B Tech and M Tech in Computer Science and Engineering form Calicut University and Dr. MGR University in the year 1999 and 2007 respectively. She is associated with Rajagiri School of Engineering & Technology, Cochin, India as an Assistant professor in Department of Information Technology. She has around 10 years of academic experience. Currently she is perusing PhD at Cochin University of Science & Technology. Her area of research is optimizing routing in Wireless Networks.

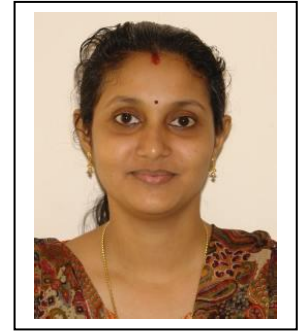